\documentclass[preprint,12pt]{elsarticle}

\usepackage[utf8]{inputenc}
\usepackage{url}
\usepackage[bookmarks=false]{hyperref}
\usepackage{etex}
 \usepackage[table]{xcolor}
\usepackage{booktabs}
\usepackage[utf8]{inputenc}
\usepackage[ruled,vlined]{algorithm2e}
\usepackage[T1]{fontenc}
\usepackage{microtype}
\usepackage{graphicx}
\usepackage{amsmath}
\usepackage{paralist}
\usepackage{tabularx}
\usepackage{balance}
\usepackage{multicol}
\usepackage{multirow}
\usepackage{pbox}
\usepackage{enumitem}
\usepackage{amssymb}
\usepackage{colortbl}
\usepackage{pifont}
\usepackage{xspace}
\usepackage{url}
\usepackage{tikz}
\usepackage{rotating}
\usepackage{float}
\usepackage{comment}
\usepackage{amssymb}
\usepackage{balance}
\usepackage[most]{tcolorbox}
\usepackage{url}
\usepackage{booktabs}
\usepackage{tikz}
\usepackage{ulem}

 \NewDocumentCommand\revised{+m}{{\color{black}#1}}
 \NewDocumentCommand\minor{+m}{{\color{black}#1}}

\begin{document}

\begin{frontmatter}

\title{Predicting Issue Types on GitHub}

\author[1]{Rafael Kallis}
\ead{rk@rafaelkallis.com}

\author[2]{Andrea Di Sorbo}
\ead{disorbo@unisannio.it}

\author[2]{Gerardo Canfora}
\ead{canfora@unisannio.it}

\author[3]{Sebastiano Panichella}
\ead{panc@zhaw.ch}

\address[1]{Valdon Group, Seilergraben 53, 8001 Zurich, Switzerland}
\address[2]{University of Sannio, Piazza Guerrazzi, 82100 Benevento, Italy}
\address[3]{Zurich University of Applied Sciences, Obere Kirchgasse 2, 8400 Winterthur, Switzerland}


\begin{abstract}
\fontsize{10}{12}\selectfont
Software maintenance and evolution involves critical activities for the success of software projects. To support such activities and keep code up-to-date and error-free, software communities make use of issue trackers, i.e., tools for signaling, handling, and addressing the issues occurring in software systems.
However, in popular projects, tens or hundreds of issue reports are daily submitted. In this context, identifying the type of each submitted report (e.g., bug report, feature request, etc.) would facilitate the management and the prioritization of the issues to address. 
To support issue handling activities, in this paper, we propose \textsc{Ticket Tagger}, a GitHub app analyzing the issue title and description through machine learning techniques to automatically recognize the types of reports submitted on GitHub and assign labels to each issue accordingly. 
We empirically evaluated the tool's prediction performance on about 30,000 GitHub issues. Our results show that the Ticket Tagger can identify the correct labels to assign to GitHub issues with reasonably high effectiveness. Considering these results and the fact that the tool is designed  to be easily integrated in the GitHub issue management process, Ticket Tagger consists in a useful solution for developers.  



\end{abstract}


\end{frontmatter}


\section{Introduction}\label{sec:intro}

Software maintenance involves tasks for mitigating potential defects in the code, as well as for evolving it according to the users' emerging needs~\cite{smr.2316}. Thus, it is crucial for the success of software projects. Issue tracking systems are tools to support these tasks by providing facilities to efficiently signal, manage, and address tickets or potential problems arising in software systems. In this context, software developers are required to timely react to issues reported in issue trackers and solve such issues by investing the lowest possible effort, to keep the costs related to software maintenance low~\cite{floris2010}. However, especially in popular projects, tens or hundreds of issues are reported daily. This complicates the issues management activities, resulting in heavier workloads for developers~\cite{bissyande2013got,PanichellaBPCA14}.

In projects hosted on GitHub, issue submitters report new issues by simply providing a title and an optional description of the issue. As issues of different types (e.g., asking questions, proposing features, signaling bugs) and quality could be submitted, GitHub also offers a customizable labeling system that can be used by developers to tag issue reports (e.g., by specifying the issue category or the related development tasks). Such labeling has positive effects on issues processing~\cite{exploring:2018}, making it easier for their management and prioritization~\cite{DBLP:conf/wcre/IzquierdoCRBC15}.
More specifically, labels assigned to issues help to classify and filter the reports, allowing more efficient issue handling processes. However, the manual labeling of issues may be labor-intensive, error-prone and time-consuming for project managers~\cite{DBLP:conf/esem/FanYYWW17} and, for this reason, labels are barely used on GitHub~\cite{exploring:2015, bissyande2013got}. 

To help maintainers dealing with issue processing, we developed \textsc{Ticket Tagger}~\cite{DBLP:conf/icsm/KallisSCP19}, a tool able to automatically label issue reports.
Differently from previous approaches aimed at automatically identifying issue types~\cite{DBLP:conf/cascon/AntoniolAPKG08, DBLP:journals/smr/ZhouTGG16}, since GitHub (according to its lightweight structure) does not provide any structured information about such issues, our tool exclusively relies on the textual features contained in the titles and descriptions of the reports to enable the automated labeling of them, immediately after they are submitted. This is beneficial for developers interested to handle new issues~\cite{DBLP:conf/wcre/IzquierdoCRBC15}.

In this paper we briefly illustrate Ticket Tagger, a GitHub app that can easily work on any software repository hosted on GitHub and 
automatically marks new issues submitted to target repositories with a relevant label. Besides, we assess the classification performance achieved by using different machine learning strategies and investigate the extent to which confounding factors of different types can degrade classification results.           

\textbf{Paper structure.} The paper is organized as follows: Section~\ref{section:ticket-classification} describes Ticket Tagger's approach and briefly presents the tool's main features, while in Section~\ref{sec:experimental-evaluation} we assess the tool's classification performance. \revised{Section \ref{sec:threats} discusses the threats that could affect the validity of our work and,  finally,} Section \ref{sec:conclusion} concludes the paper outlining future research directions.

\section{Approach and Tool's Overview}\label{section:ticket-classification}

To classify an issue report, Ticket Tagger processes the report's title and body to represent the textual information (extracted from the issue) in a vectorial space. By inspecting the resulting components, the tool can assign a relevant label to the mentioned report.

\textbf{The Machine Learning Model}. Different Machine Learning (ML) algorithms can be adopted to efficiently classify textual information~\cite{joulin2016bag,SorboPVPCG16,PanichellaSGVCG15}. However,
complex ML strategies may require a long time for training and consume a lot of memory. Since we wanted to deploy the model on low-end server hardware\footnote{\fontsize{8}{10}\selectfont AWS EC2 \texttt{t2.nano} (1 vCPU, 512 MB RAM, 20 GB SSD)}, we opted for \textit{fastText}, a tool using linear models with a rank constraint and fast loss approximation, able to achieve comparable classification results to several deep learning-based approaches~\cite{joulin2016bag}.

\textbf{Issue Reports pre-processing and Vectorial Representation}.
For allowing the fastText linear classifier to make issue type predictions, the title and body of the reports are concatenated into a single textual paragraph. The resulting text is then tokenized and the tokenized text represents the source for obtaining the bag of words representation of the issue. This bag of words representation, in which each word is represented by a vector of \textit{character n-grams}, is the input of the fastText based classifier.



\textbf{Issues Classification}. The fastText model classifies issues by minimizing the following objective function over $N$ possible labels:
\vspace{-1.5mm}

\begin{equation*}
    -\frac{1}{N}\sum_{n=1}^{N} y_n \log (f(B A x_n))
\end{equation*}

where $x_n$ is a bag of features, $A$ represents the weight dictionary of the average text embeddings, $B$ is the weight dictionary that converts the embedding to pre-softmax values for each class, and $f$ is the hierarchical softmax function used to minimize computational complexity~\cite{DBLP:conf/icsm/KallisSCP19}.

\begin{figure*}[t!]
\vspace{-10mm}
    \centering
    \includegraphics[scale=0.32]{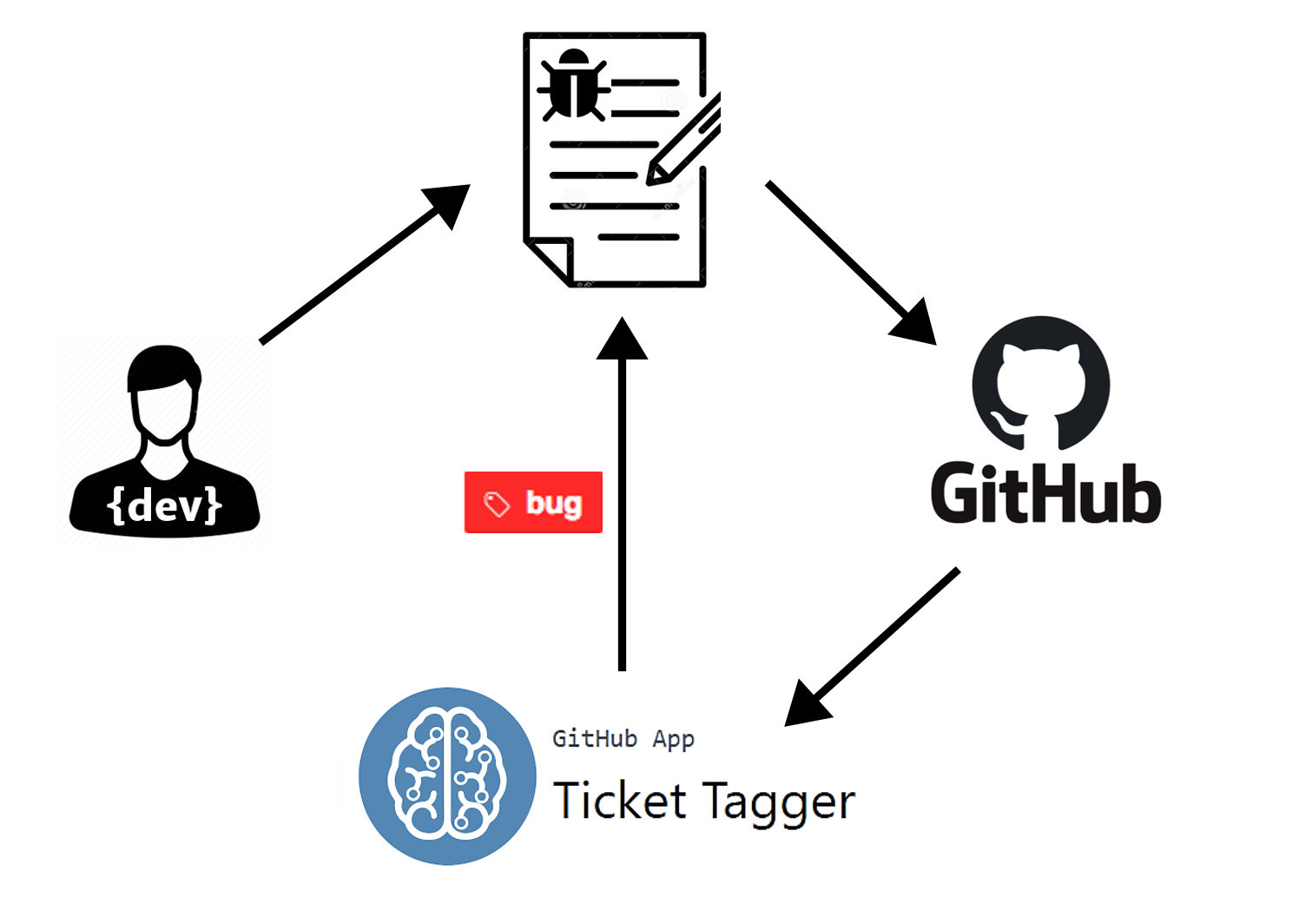}
\vspace{-6mm}
    \caption{Ticket Tagger issue labeling process.}
    \label{fig:process}
    \vspace{-4mm}
\end{figure*}

We set fastText by using the default values for most of the parameters\footnote{\fontsize{8}{10}\selectfont For further details, see  \url{https://fasttext.cc/docs/en/options.html}} and applied the following customization:
\begin{itemize}
\vspace{-1.5mm}
    \item word n-gram features are not captured, i.e., \texttt{wordNgrams} parameter;
    \vspace{-2mm}
    \item we only consider words that occur at least 14 times in the dataset ,i.e., \texttt{minCount} parameter.
    \vspace{-1.5mm}
\end{itemize}
Both settings have been applied according to the disk constraints of our server hardware. Indeed, these decisions allowed us to obtain a trained model requiring less than 5 MB of disk space whilst only imposing a \textless 10\% performance penalty.

Ticket Tagger is currently able to classify issues according to three categories reflecting the intent~\cite{SorboPVPCG16, DBLP:conf/sigsoft/SorboPASVCG16} of the writer: \textit{bug report}, \textit{enhancement}, and \textit{question}.
These labels are included by default in every GitHub repository and they are the three labels most used on GitHub~\cite{exploring:2015}. 
Obviously, our model \minor{is designed to be easily re-trained} to adapt Ticket Tagger to specific projects' needs, enabling the prediction of additional issue types.  


\textbf{Tool's Overview}. When a new issue report is submitted to a GitHub repository on which Ticket Tagger is installed, the tool automatically assigns a relevant label to the new report. In particular, Ticket Tagger is a Node.js-based GitHub app, that automatically (i) gathers issue reports information from a GitHub repository, and (ii) labels the newly reported issues, by leveraging the pre-trained fastText model previously discussed. 
The app is freely accessible and can be easily installed onto any existing GitHub repositories. By navigating to the Ticket Tagger app webpage\footnote{\fontsize{8}{10}\selectfont \url{https://github.com/apps/ticket-tagger}}, 
to install Ticket Tagger on a target repository, the repository administrator has to click on the ``Install'' button, specify the repository, and that's it. From this moment on, as depicted in Figure \ref{fig:process}, when a user opens a new issue ticket on the repository, GitHub calls the hook endpoint exposed by Ticket Tagger and references the information related to the newly created issue. Such information is used by the app to classify the ticket. In order to automatically label the issue report, GitHub provides a temporary access token to Ticket Tagger, which is consumed by assigning the predicted label to the issue. 
The automated issue labeling performed by Ticket Tagger allows the developers to (i) timely react to urgent issues, (ii) postpone less impelling tasks (such as enhancement requests), or (iii) assign the questions to specific users.

\section{Performance Evaluation}
\label{sec:experimental-evaluation}




\vspace{-2mm}
In this section, we describe the \minor{datasets} and baseline approach used to  assess the classification performance of the fastText model integrated into Ticket Tagger (described in Section~\ref{section:ticket-classification}).  

\minor{\textbf{Datasets Construction}. 
For assessing Ticket Tagger's effectiveness in classifying GitHub issues we collected two datasets. The first dataset, $D_{balanced}$, contains 30,000 issues}\footnote{ \fontsize{8}{10}\selectfont \url{https://tinyurl.com/y23kgdro}}. This dataset was obtained by first collecting issues from 12,112 heterogeneous projects, this by querying the GitHub Archive\footnote{\fontsize{8}{10}\selectfont \url{https://gharchive.org}} using Google BigQuery\footnote{ \fontsize{8}{10}\selectfont \url{https://cloud.google.com/bigquery}}. After this initial step, we randomly sampled issues from the set of all GitHub issues closed during \revised{February} 2018, thus selected all issues having label matching the following strings: \textit{bug}, \textit{enhancement} or \textit{question}.
With this random selection process, we selected, on average, $2.48$ issues for each project (median $=1$ and standard deviation $=15.78$). 
One third of the 30,000 issues had the bug label assigned; one third issues had the enhancement label\footnote{\fontsize{8}{10}\selectfont This label refers to improvements and new features.} assigned; while the remaining 10,000 issues had the question label assigned.
\minor{To construct the second dataset, $D_{unbalanced}$, we ran a query\footnote{\scriptsize\url{https://tickettagger.blob.core.windows.net/scripts/github-labels-top3-34k.sql}} over the GitHub Archive using Google BigQuery. We queried for issues containing any of the three labels, i.e., bug, enhancement and question, between the 1st and 9th of March 2018 in the GitHub Archive, 
obtaining approximately 34,000 issues\footnote{\scriptsize\url{https://tickettagger.blob.core.windows.net/datasets/github-labels-top3-34k.csv}}. The resulting distribution of issue types in $D_{unbalanced}$ is as follows: 16,355 (48\%) tickets labeled as bug, 14,228 (41.8\%) tickets marked as enhancement, and 3,458 (10.2\%) question issues.}
\minor{While the first dataset, $D_{balanced}$, contains an identical number of tickets from each category, the second one, $D_{unbalanced}$, presents an unbalanced distribution of labels and is more representative of reality.}



\textbf{Evaluation Methodology}.  
\minor{The goal of our experiments is twofold. On the one hand we compare Ticker Tagger against a baseline approach, 
to observe whether more simple ML-based approaches are able to achieve comparable or better results than  Ticker Tagger}.  
\minor{On the other hand, we evaluate the extent to which Ticket Tagger is able to automatically identify the correct labels to assign to issue reports in a realistic scenario.}
\minor{More specifically, we compare Ticket Tagger with 
the J48 machine learning (ML) algorithm that was successfully used in previous work concerning the assessment of ML strategies for textual classification problems \cite{PanichellaSGVCG15, DBLP:conf/sigsoft/SorboPASVCG16}.
To perform such a comparison, a 10-fold cross validation strategy \cite{10489157} on $D_{balanced}$ is used for evaluating the classification performance achieved by both Ticket Tagger and the baseline J48 ML algorithm.}

\minor{For training the J48 model, we leverage all the terms contained in both titles and descriptions of issues in our dataset to build a document-term matrix $M$, where each row represents an issue of our dataset, and each 
column represents a term. Every entry $M_{ij}$ of the aforementioned matrix represents the weight or importance of the \textit{j-th} term in the \textit{i-th} issue, computed according to the \textit{tf-idf} weighting scheme~\cite{BaezaYates:1999} that has been successfully used in recent work concerning the classification of GitHub issues~\cite{DBLP:conf/esem/FanYYWW17} and vulnerabilities~\cite{DBLP:journals/jss/RussoSVC19}.}
\revised{It is worth noticing that, for ensuring a fair comparison between the two models, in applying J48, we do not perform any model tuning and pre-processing of the data, since also fastText is used in the same way. In the future, we are interested in investigating the pre-processing steps and parameters tuning required to achieve better results.}
\revised{Furthermore,} the evaluation is performed without the custom settings used for reducing fastText's disk space (described in Section~\ref{section:ticket-classification}).

\minor{With the aim of assessing the Ticket Tagger's capability of recognizing issue types in a realistic setting, i.e., unbalanced distribution of issue types, we carry out a further experiment in which  Ticket Tagger is trained on the whole balanced dataset, $D_{balanced}$, and the unbalanced dataset, $D_{unbalanced}$, is used for evaluating the classification performance.
This particular setting , i.e., balanced training set and unbalanced test set, is motivated by the need to avoid that the resulting model is biased towards the majority class(es).
Well-known information retrieval metrics, namely  precision, recall, and F-measure~\cite{BaezaYates:1999}, are adopted to evaluate the classification performance in our experiments. 
}

\begin{table}[]
    \centering
    \tiny
    \caption{\minor{Precision, Recall and F-measure of \textit{bug}, \textit{enhancement} and \textit{question} labels for Ticker Tagger and the baseline J48 model, obtained using a 10-fold cross validation over $D_{balanced}$. Differences with the baseline approach are highlighted in \textbf{bold}.}}
    \fontsize{8}{9}\selectfont
    \label{table:model}
    {\color{black}
    \begin{tabular}{l|c|c c c c}
    \toprule
    Approach &  \bf Metrics & \bf Bug & \bf Enhancement & \bf Question \\
    \midrule
    & Precision & 0.82 \bf(+0.24) & 0.89 \bf(+0.29)& 0.78 \bf(+0.13)\\
   \it Ticker Tagger & Recall & 0.84 \bf(+0.25) & 0.76 \bf (+0.13) & 0.87 \bf (+0.26)\\
    & F-measure & 0.83 \bf(+0.24) & 0.82 \bf(+0.20) & 0.83 \bf (+0.20)\\
    \hline
        & Precision & 0.58 & 0.60 & 0.65\\
   \it J48 & Recall & 0.59 & 0.63 & 0.61\\
    & F-measure & 0.59 & 0.62 & 0.63\\

    \bottomrule
    \end{tabular}
    }
\end{table}

\textbf{Results}. 
\minor{Table~\ref{table:model} reports the classification performance achieved by both Ticket Tagger and the baseline approach (J48) using 10-fold cross validation on $D_{balanced}$.
In particular, Table~\ref{table:model} shows how Ticker Tagger obtained F-measure values above 0.80 for each considered label, confirming the practical usefulness of the proposed approach for improving the issue management practices on GitHub.  In addition, we can observe how for $D_{balanced}$, Ticket Tagger always outperforms the baseline approach (J48) for all labels and in all precision, recall, and F-measure metrics.} 

\minor{Table~\ref{table:model-unbalanced} shows the performance of Ticket Tagger in identifying bug, enhancement and question issues, when trained on $D_{balanced}$ and tested on $D_{unbalanced}$. The results of this second experiment highlight that our tool automatically identifies issues of the bug and enhancement types with reasonably high effectiveness, i.e., F-measure of about 0.75, while lower classification performance is obtained for the question category. 
On the one hand, these findings confirm the practical usefulness of our tool, as it achieves reasonably high performance in automatically recognizing issues reporting bugs or requesting features. These are the most important feedback for developers interested in performing software maintenance and evolution activities
~\cite{DBLP:conf/sigsoft/SorboPASVCG16}. On the other hand, we believe that further efforts and tunings are required to improve the tool's capability of recognizing issues of the question type.}  

\begin{table}[h]
    \centering
        \tiny
    \caption{\minor{Precision, Recall and F-measure of \textit{bug}, \textit{enhancement} and \textit{question} labels, when Ticker Tagger is trained on $D_{balanced}$ and tested on $D_{unbalanced}$. The proportion of tickets are $48$\%, $41.8$\% and $10.2$\%, respectively.}}
\revised{
    \fontsize{8}{9}\selectfont
    \label{table:model-unbalanced}
    {\color{black}
    \begin{tabular}{l|c c c}
    \toprule
    \bf Metrics & \bf Bug & \bf Enhancement & \bf Question \\
    \midrule
    Precision & $0.79$ & $0.73$ & $0.44$ \\
    Recall & $0.72$ & $0.74$ & $0.53$ \\
    F-measure & $0.75$ & $0.74$ & $0.48$ \\
    \bottomrule
    \end{tabular}
    }
}
\end{table}

In recent work, Herbold et al.~\cite{abs-2003-05357} considered Ticker Tagger in a quantitative comparison, showing that fastText outperforms the competition concerning the issue labeling  problem, this without particular tuning. 
Herbold et al.'s approach achieves slightly higher precision results than our model because it leverages the auto-tuning feature, a feature that we did not use in Ticket Tagger. Thus, such small improvements in prediction performance are due to structural information about the issues used. 

\textbf{Discussion of confounding factors.} There are several factors that can potentially influence Ticker Tagger's performance, as discussed below.
\begin{itemize}
\vspace{-2mm}
    \item[(i)] \textit{Impact of function words}: \minor{For issues belonging to the bug and enhancement classes both precision and recall are above 0.70, while Ticket Tagger produces higher numbers of false positives and false negatives for the question category, i.e., a lower precision and a lower recall are achieved for this class.
    We believe that the strong use of function words, e.g., ``how'' or ``what'' that typically introduce questions, in the issue title or description  could lead the classifier to erroneously assign the question label to issues that actually belong to  different classes and, consequently, this degrades the precision achieved for the question category. In addition, the lower recall obtained for this class could be connected with the fact that developers (and users) ask questions about a wide range of topics~\cite{DBLP:journals/jcst/YangLXWS16}, making it hard to learn all the  patterns that could lead to the assignment of this label. }
    \vspace{-2mm}
    \item[(ii)] \textit{Impact of Language Consistency in Issue Tickets}: we observe whether the ticket's language affects the performance of our model. Thus, we generated two datasets, one containing 24,600 English tickets and one baseline dataset of 24,600 tickets with random tickets sampled using the same strategy described in Section \textit{Dataset Construction}. To generate the dataset comprising 24,600 English tickets, we used a javascript port of \textit{guess language}\footnote{\scriptsize\url{https://github.com/wooorm/franc}}, a tool using heuristics based on character sets and trigrams for automatically detecting the language of the text.
    Results in Table~\ref{table:model-factors} suggest that language consistency in issue tickets has a positive effect on the classification performance.
    \vspace{-2mm}
    \item[(iii)] \textit{Presence of Code Snippets in Issue Tickets}: we observe whether the presence of code snippets in tickets affects the performance of our model. Thus, we generated two datasets, one characterized by 6,000 tickets containing code snippets and one baseline dataset of 6,000 tickets sampled at random using the previously mentioned method. 
    In particular, the presence of code snippets is recognized by detecting pieces of text enclosed in triple backticks, which is the special syntax recommended by the \textit{GitHub Flavored Markdown} language\footnote{\scriptsize\url{https://docs.github.com/en/github/writing-on-github/basic-writing-and-formatting-syntax}} to highlight code snippets.
    Results in Table~\ref{table:model-factors} show that the presence of snippets 
    \revised{does not significantly impact classification performance.}
\end{itemize}

\begin{table}[]
    \centering
        \tiny
            \vspace{-4mm}
    \caption{Precision, Recall and F-measure of \textit{bug}, \textit{enhancement} and \textit{question} labels for Ticker Tagger are computed using a 10-fold cross validation. Only differences among the various treatments (CONSISTENT LANGUAGE and CODE SNIPPET PRESENCE) are reported for the sake of brevity.}
\revised{
    \fontsize{8}{9}\selectfont
    \label{table:model-factors}
    \begin{tabular}{l|c|c c c}
    \toprule
    Approach &  \bf Metrics & \bf Bug & \bf Enhancement & \bf Question \\
    \midrule
    \multirow{3}{8em}{\it CONSISTENT LANGUAGE} & Precision & $-2.5$\% & $+6.3$\% & $+2.0$\% \\
    & Recall & $+9.4$\% & $-1.8$\% & $+1.7$\% \\
    & F-measure & $+4.0$\% & $+2.4$\% & $+1.9$\% \\
    \hline
    \multirow{3}{8em}{\it CODE SNIPPET PRESENCE} & Precision & $+0.6$\% & $-2.0$\% & $-0.4$\% \\
    & Recall & $-3.1$\% & $+2.5$\% & $+0.5$\% \\
    & F-measure & $-0.3$\% & $+0.1$\% & $+0.0$\% \\

   \bottomrule
    \end{tabular}
    
}
\end{table}

\section{Threats to Validity}\label{sec:threats}

\textbf{Threats to construct validity}.
We compared Ticket Tagger with a baseline approach (J48) on a dataset comprising equal numbers of bugs, enhancements and questions. This could represent a threat to construct validity as in real scenarios the distributions of the different types of issues may be unbalanced. To counteract this issue, we also assessed Ticket Tagger on a second unbalanced dataset where the proportion between the different classes is close to reality.

\textbf{Threats to internal validity}. 
Our results could be misleading if a significant percentage of collected issues would be subject to re-labeling. To mitigate this concern and reduce the likelihood of re-labeling for the considered samples, we collected GitHub issues having the \textit{closed} status assigned. 

\textbf{Threats to external validity}. 
The main threat to external validity is related to the potential specificity of our datasets. The collected issues could not be adequately representative of all the issues present on GitHub. However, to increase the heterogeneity of data, we selected issues from projects (i) having different natures, (ii) implemented through different programming languages,  and (iii) developed by different developers' communities. To further confirm the low specificity of our datasets and the quality of our results, in recent work Ticker Tagger was considered in a quantitative comparison~\cite{abs-2003-05357}, which demonstrated that fastText outperforms state-of-the-art approaches addressing the issue labeling problem. 

\section{Conclusion}\label{sec:conclusion}

In this work, we presented Ticket Tagger, an app that we released  on the GitHub marketplace, that automatically assigns suitable labels to issues opened on GitHub projects. 
The core of Ticket Tagger is represented by a machine learning model that analyzes the title and the textual description of issues in order to determine whether such an issue can be labeled as a  bug report, a feature request or a question.

With the aim of assessing the classification performance achieved by our tool, we conducted four main evaluation experiments. 
The results of such evaluation showed that  Ticket Tagger allows to automatically assign labels with reasonably high levels of precision and recall, outperforming results of a baseline approach. Our findings have also shown that the use of a consistent language can improve Ticket Tagger classification performance, while the presence of code snippets does not affect the results significantly.

Future work will be aimed (i) at comparing Ticket Tagger's accuracy and functionality with other existing solutions, as well as (ii) at investigating its usefulness through the analysis of direct feedback from end-users.

\section{Acknowledgements}

The authors express their gratitude and appreciation towards the anonymous reviewers who dedicated their considerable time and expertise to the paper. Sebastiano Panichella gratefully acknowledges the Horizon 2020 (EU Commission) support for the
project \textit{COSMOS} (DevOps for Complex Cyber-physical Systems), Project No. 957254-COSMOS).

{
\fontsize{7}{8}\selectfont
\bibliographystyle{elsarticle-num}
\bibliography{tickettagger}

\begin{thebibliography}{10}
\expandafter\ifx\csname url\endcsname\relax
  \def\url#1{\texttt{#1}}\fi
\expandafter\ifx\csname urlprefix\endcsname\relax\def\urlprefix{URL }\fi
\expandafter\ifx\csname href\endcsname\relax
  \def\href#1#2{#2} \def\path#1{#1}\fi

\bibitem{smr.2316}
A.~Di~Sorbo, G.~Grano, C.~Aaron~Visaggio, S.~Panichella,
  \href{https://onlinelibrary.wiley.com/doi/pdf/10.1002/smr.2316}{Investigating
  the criticality of user-reported issues through their relations with app
  rating}, Journal of Software: Evolution and Process n/a~(n/a)  e2316.
\newline\urlprefix\url{https://onlinelibrary.wiley.com/doi/pdf/10.1002/smr.2316}

\bibitem{floris2010}
P.~Floris, H.~Vogt~Harald, How to save on software maintenance costs, omnext
  white paper, SOURCE 2 VALUE.

\bibitem{bissyande2013got}
T.~F. Bissyand{\'e}, D.~Lo, L.~Jiang, L.~R{\'e}veillere, J.~Klein, Y.~Le~Traon,
  Got issues? who cares about it? a large scale investigation of issue trackers
  from github, in: International symposium on software reliability engineering,
  2013, pp. 188--197.

\bibitem{PanichellaBPCA14}
S.~Panichella, G.~Bavota, M.~D. Penta, G.~Canfora, G.~Antoniol, How developers'
  collaborations identified from different sources tell us about code changes,
  in: {ICSME}, {IEEE}, 2014, pp. 251--260.

\bibitem{exploring:2018}
Z.~Liao, D.~He, Z.~Chen, X.~Fan, Y.~Zhang, S.~Liu, Exploring the
  characteristics of issue-related behaviors in github using visualization
  techniques, IEEE Access 6 (2018) 24003--24015.

\bibitem{DBLP:conf/wcre/IzquierdoCRBC15}
J.~L.~C. Izquierdo, V.~Cosentino, B.~Rolandi, A.~Bergel, J.~Cabot, Gila: Github
  label analyzer, in: International Conference on Software Analysis, Evolution,
  and Reengineering, {SANER}, 2015, pp. 479--483.

\bibitem{DBLP:conf/esem/FanYYWW17}
Q.~Fan, Y.~Yu, G.~Yin, T.~Wang, H.~Wang, Where is the road for issue reports
  classification based on text mining?, in: International Symposium on
  Empirical Software Engineering and Measurement, {ESEM} 2017, 2017, pp.
  121--130.

\bibitem{exploring:2015}
J.~Cabot, J.~L.~C. Izquierdo, V.~Cosentino, B.~Rolandi, Exploring the use of
  labels to categorize issues in open-source software projects, in:
  International Conference on Software Analysis, Evolution, and Reengineering
  (SANER), 2015, pp. 550--554.

\bibitem{DBLP:conf/icsm/KallisSCP19}
R.~Kallis, A.~{Di Sorbo}, G.~Canfora, S.~Panichella, Ticket tagger: Machine
  learning driven issue classification, in: International Conference on
  Software Maintenance and Evolution, {ICSME}, 2019, pp. 406--409.

\bibitem{DBLP:conf/cascon/AntoniolAPKG08}
G.~Antoniol, K.~Ayari, M.~Di~Penta, F.~Khomh, Y.~Gu{\'{e}}h{\'{e}}neuc, Is it a
  bug or an enhancement?: a text-based approach to classify change requests,
  in: Conference of the Centre for Advanced Studies on Collaborative Research,
  2008, p.~23.

\bibitem{DBLP:journals/smr/ZhouTGG16}
Y.~Zhou, Y.~Tong, R.~Gu, H.~C. Gall, Combining text mining and data mining for
  bug report classification, Journal of Software: Evolution and Process 28~(3)
  (2016) 150--176.

\bibitem{joulin2016bag}
A.~Joulin, E.~Grave, P.~Bojanowski, T.~Mikolov, Bag of tricks for efficient
  text classification, in: Proceedings of the 15th Conference of the European
  Chapter of the Association for Computational Linguistics, {EACL} 2017,
  Valencia, Spain, April 3-7, 2017, Volume 2: Short Papers, 2017, pp. 427--431.

\bibitem{SorboPVPCG16}
A.~Di~Sorbo, S.~Panichella, C.~A. Visaggio, M.~Di~Penta, G.~Canfora, H.~C.
  Gall, {DECA:} development emails content analyzer, in: International
  Conference on Software Engineering, {ICSE} 2016 - Companion Volume, 2016, pp.
  641--644.

\bibitem{PanichellaSGVCG15}
S.~Panichella, A.~Di~Sorbo, E.~Guzman, C.~A. Visaggio, G.~Canfora, H.~C. Gall,
  How can {I} improve my app? {Classifying} user reviews for software
  maintenance and evolution, in: International Conference on Software
  Maintenance and Evolution, {ICSME}, 2015, pp. 281--290.

\bibitem{DBLP:conf/sigsoft/SorboPASVCG16}
A.~{Di Sorbo}, S.~Panichella, C.~V. Alexandru, J.~Shimagaki, C.~A. Visaggio,
  G.~Canfora, H.~C. Gall, What would users change in my app? summarizing app
  reviews for recommending software changes, in: Proceedings of the 24th {ACM}
  {SIGSOFT} International Symposium on Foundations of Software Engineering,
  {FSE} 2016, 2016, pp. 499--510.

\bibitem{10489157}
D.~M. Allen, The relationship between variable selection and data agumentation
  and a method for prediction, Technometrics 16~(1) (1974) 125--127.
\newblock \href {http://dx.doi.org/10.1080/00401706.1974.10489157}
  {\path{doi:10.1080/00401706.1974.10489157}}.

\bibitem{BaezaYates:1999}
R.~A. Baeza-Yates, B.~Ribeiro-Neto, Modern Information Retrieval,
  Addison-Wesley Longman Publishing Co., Inc., Boston, MA, USA, 1999.

\bibitem{DBLP:journals/jss/RussoSVC19}
E.~R. Russo, A.~D. Sorbo, C.~A. Visaggio, G.~Canfora, Summarizing
  vulnerabilities' descriptions to support experts during vulnerability
  assessment activities, J. Syst. Softw. 156 (2019) 84--99.

\bibitem{abs-2003-05357}
S.~Herbold, A.~Trautsch, F.~Trautsch, On the feasibility of automated
  prediction of bug and non-bug issues, Empir. Softw. Eng. 25~(6) (2020)
  5333--5369.

\bibitem{DBLP:journals/jcst/YangLXWS16}
X.~Yang, D.~Lo, X.~Xia, Z.~Wan, J.~Sun, What security questions do developers
  ask? {A} large-scale study of stack overflow posts, J. Comput. Sci. Technol.
  31~(5) (2016) 910--924.

\end{thebibliography}
}

\clearpage

\section*{Appendix}
\label{sec:appendix}

\subsection*{Current executable software version}

\begin{table}[!h]
\begin{tabular}{|l|p{6.5cm}|p{6.5cm}|}
\hline
\textbf{Nr.} & \multicolumn{2}{l|}{\textbf{Software metadata description}} \\
\hline
S1 & Current software version & 2.1.4 \\
\hline
S2 & Permanent link to executables of this version &  \url{https://github.com/rafaelkallis/ticket-tagger/releases/tag/v2.1.4} \\
\hline
S3 & Legal Software License   & GNU General Public License (GPL) \\
\hline
S4 & Computing platform / Operating System & macOS, Linux \\
\hline
S5 & Installation requirements \& dependencies & nodejs 12 \\
\hline
S6 & If available Link to user manual - if formally published include a reference to the publication in the reference list & https://github.com/rafaelkallis/ticket-tagger/blob/master/README.md \\
\hline
S6 & Support email for questions & \url{rk@rafaelkallis.com} \\
\hline
\end{tabular}
\caption{Software metadata}
\label{} 
\end{table}

\clearpage

\subsection*{Current code version}
\label{sec:current-code-version}

\begin{table}[!h]
\begin{tabular}{|l|p{6.5cm}|p{6.5cm}|}
\hline
\textbf{Nr.} & \multicolumn{2}{l|}{\textbf{Code metadata description}} \\
\hline
C1 & Current code version & 2.1.4 \\
\hline
C2 & Permanent link to code/repository used of this code version &  \url{https://github.com/rafaelkallis/ticket-tagger} \\
\hline
C3 & Legal Code License   & GNU General Public License (GPL) \\
\hline
C4 & Code versioning system used & git \\
\hline
C5 & Software code languages, tools, and services used & javascript, nodejs, heroku, fasttext \\
\hline
C6 & Compilation requirements, operating environments \& dependencies & nodejs 12 \\
\hline
C7 & If available Link to developer documentation/manual & \url{https://github.com/rafaelkallis/ticket-tagger} \\
\hline
C8 & Support email for questions & \url{rk@rafaelkallis.com} \\
\hline
\end{tabular}
\caption{Code metadata}
\label{} 
\end{table}

\end{document}